\documentclass[a4paper,12pt]{article}

\usepackage{a4}
\usepackage{latexsym, amssymb}

\usepackage{graphicx}

\def\thefootnote{\fnsymbol{footnote}}

\begin{document}

\begin{titlepage}

\begin{flushright}
{\small 
December 17, 2013}  
\end{flushright}

\vspace*{0.5cm}
\begin{center}
{\Large {\bf 
Status of hadronic light-by-light scattering \\[0.2cm]
in the muon $g-2$~\footnote{Based on presentations at the {\it International
  Workshop on $e^+ e^-$ collisions from $\Phi$ to 
     $\Psi$ 2013 (PHIPSI13)}, September 9-12, 2013, Rome, Italy, 
  the 14th Meeting of the {\it Working Group on Radiative Corrections
    and Monte Carlo Generators for Low Energies}, September 13, 2013, Frascati,
  Italy, and the {\it LC13 Workshop: Exploring QCD from the infrared
    regime to heavy flavour scales at B-factories, the LHC and a
    Linear Collider}, September 16-20, 2013, Trento, Italy. Some
  material added compared to the text submitted to the proceedings of
  the LC13 Workshop.}}} \\[1cm]
Andreas Nyf\/feler\footnote{nyf\/feler@hri.res.in} \\[0.5cm]
{\it Regional Centre for Accelerator-based Particle Physics, 
Harish-Chandra Research Institute, Chhatnag Road, Jhusi,  
Allahabad - 211 019, India} \\[0.3cm]
{\it Albert Einstein Center for Fundamental Physics, Institute for
  Theoretical Physics, University of Bern, Sidlerstrasse 5, CH-3012
  Bern, Switzerland}
\end{center}

\vspace*{0.5cm}
\begin{abstract}
We give an update on the status of the hadronic light-by-light
scattering contribution to the muon $g-2$.  We review recent work by
various groups, list some of the open problems and give an outlook on
how to better control the uncertainty of this contribution. This is
necessary in order to fully profit from planned future muon $g-2$
experiments to test the Standard Model. Despite some recent
developments, we think that the estimate $a_\mu^{\rm HLbL} = (116 \pm
40) \times 10^{-11}$ still gives a fair description of the current
situation.
\end{abstract}

\vspace*{0.5cm}

\end{titlepage}


\renewcommand{\thefootnote}{\arabic{footnote}}
\setcounter{footnote}{0}

\section{Introduction}

The anomalous magnetic moment of the muon has served over many years
as an important test of the Standard Model, see the
reviews~\cite{JN_09,Miller_etal_12}. It is also sensitive to potential
contributions from New Physics. The current status of the muon $g-2$
is summarized in Table~\ref{tab:status_g-2} where we list the
different contributions in theory (QED, weak, hadronic) from various
recent sources and compare with the experimental value. More
references to earlier work can be found in the quoted papers and in
Refs.~\cite{JN_09,Miller_etal_12}. The experimental world average is
dominated by the final result of the Brookhaven muon $g-2$
experiment~\cite{BNL_g-2}, corrected for a small shift in the ratio of
the magnetic moments of the muon and the proton~\cite{CODATA_2008}. We
observe a difference between experiment and theory of more than three
standard deviations
\begin{equation}
a_\mu^{\rm exp} - a_\mu^{\rm th} = (293 \pm 88) \times 10^{-11} 
\qquad \quad (3.3\sigma).  
\end{equation}
Unfortunately, the theoretical
uncertainties~\cite{JN_09,Miller_etal_12,Venanzoni_talk_Blum_et_al_13}
from hadronic vacuum polarization (HVP) and hadronic light-by-light
scattering (HLbL) make it difficult to interpret this discrepancy as a
clear sign of New Physics. Most recent
evaluations~\cite{JN_09,JS_11,recent_evaluations,Miller_etal_12,Venanzoni_talk_Blum_et_al_13},
which differ slightly in the treatment of the hadronic contributions,
obtain deviations of $3-4\sigma$. In Ref.~\cite{Benayoun_etal} the
hadronic cross-section data below $1~\mbox{GeV}$ was fitted to a
(broken) Hidden Local Symmetry (HLS) model and a discrepancy in the
muon $g-2$ between $3.7-4.9\sigma$ was observed, depending on the
selected data.

\begin{table}[h]
  \caption{Standard Model contributions to  $a_\mu \times 10^{11}$ and
    comparison of theory and experiment.}    
  \label{tab:status_g-2}
  \begin{center}
  \begin{tabular}{|l|r|r|c|}
    \hline 
    Contribution & \multicolumn{1}{|c|}{~~~~~Value~~~~~} &
    \multicolumn{1}{|c|}{~~~Error~~~~}  & ~~Reference~~ \\   
    \hline 
    QED               & 116~584~718.853            &  0.036 &
    \cite{Aoyama_etal_12} \\ 
    Weak              &         153.6\hphantom{00} &  1.0\hphantom{00}
    & \cite{GSS_13} \\ 
    Leading order HVP &       6~907.5\hphantom{00} & 47.2\hphantom{00}
    & \cite{JS_11} \\ 
    Higher order HVP  &        -100.3\hphantom{00} &  2.2\hphantom{00}
    & \cite{JS_11} \\ 
    HLbL              &         116\hphantom{.000} & 40\hphantom{.000}
    & \cite{N_09,JN_09} \\   
    Theory (total)    & 116~591~796\hphantom{.000} & 62\hphantom{.000}
    & - \\     
    \hline 
    Experiment        & 116~592~089\hphantom{.000} & 63\hphantom{.000}
    & \cite{BNL_g-2} \\
    Experiment - Theory ($3.3\sigma$) &  293\hphantom{.000} &
    88\hphantom{.000} & - \\  
    \hline 
  \end{tabular}
  \end{center} 
\end{table}

\begin{figure}[t]
\includegraphics[width=\textwidth]{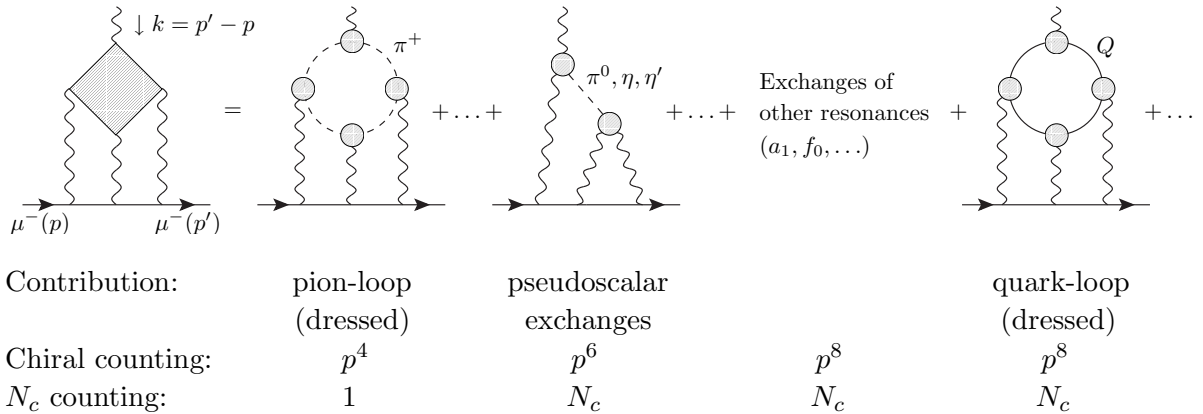}    

\vspace*{0.2cm}
{\small 
\hspace*{0.32cm}Contribution: \hspace*{1.2cm} pion-loop  \hspace*{1cm}
pseudoscalar  \hspace*{4cm}  quark-loop \\{}  
\hspace*{4cm} (dressed) \hspace*{1.25cm} exchanges \hspace*{4.325cm}
(dressed) \\{} 
\hspace*{0.32cm}Chiral counting: \hspace*{1.4cm}  $p^4$  \hspace*{2.4cm}
$p^6$  \hspace*{2.6cm} $p^8$  \hspace*{2.35cm} $p^8$ \\{}  
\hspace*{0.32cm}$N_c$ counting:   \hspace*{2cm}  $1$   \hspace*{2.5cm}
$N_c$  \hspace*{2.5cm} $N_c$  \hspace*{2.25cm} $N_c$  
}
\caption{The different contributions to HLbL scattering and their
  chiral and large-$N_c$ counting.}  
\label{fig:HLbL}
\end{figure}

The HLbL contribution to the muon $g-2$ involves the Green function of
four electromagnetic currents, connected to off-shell photons, see
Figure~\ref{fig:HLbL} and Ref.~\cite{JN_09} for details and
references. The relevant scales for the off-shell photons in HLbL are
about $0-2~\mbox{GeV}$, i.e.\ larger than the muon mass, and therefore
a pure low-energy effective field theory approach with muons, photons
and pions fails~\cite{Knecht_etal_02}. In contrast to the HVP
contribution, HLbL cannot be directly related to experimental data and
therefore various models have been employed to estimate HLbL. One uses
some hadronic model with exchanges and loops of resonances at low
energies and some form of (dressed, constituent) quark-loop at high
energies as short-distance complement of the low-energy hadronic
models. The dependence on several momenta leads, however, to a mixing
of long and short distances and makes it difficult to avoid a double
counting of quark-gluon and hadronic contributions. In
Ref.~\cite{deRafael_94} a classification of the different
contributions to HLbL based on the chiral and large-$N_c$ counting was
proposed, see Figure~\ref{fig:HLbL}. In general, all the interactions
of the hadrons or the quarks with the photons are dressed by some form
factors, e.g.\ via $\rho-\gamma$ mixing. Note that in the Feynman
diagrams in Fig.~\ref{fig:HLbL} form factors with off-shell photons
and off-shell hadrons enter~\cite{J_essentials_07}. Constraints on the
models can be obtained from experimental data, e.g.\ on the various
form factors, and from theory, e.g.\ chiral perturbation theory at low
energies and short-distance constraints from perturbative QCD and the
Operator Product Expansion (OPE) at high momenta.

\section{Current status of HLbL and recent developments}

\begin{table}[h]
  \caption{Summary of selected estimates for the different
    contributions to $a_\mu^{\rm HLbL} \times 10^{11}$. For
    comparison, the last line shows some results when no form factors 
    are used (undressed).}  
  \label{tab:HLbL_summary}
  \begin{center} 
  \begin{tabular}{|c|c|c|c|c|c|c|}
    \hline 
    $\pi,K$-loops & $\pi^0,\eta,\eta^\prime$ & axial-vect.\ & scalars & 
    quark-loop & Total & Ref.\ \\   
    \hline 
    $-4.5(8.1)$ & 82.7(6.4) & 1.7(1.7) & - & 9.7(11.1) & 89.6(15.4) & 
    \cite{HKS} \\ 
    $-19(13)$ & 85(13) & 2.5(1.0) & $-6.8(2.0)$ & 21(3) & 83(32) &
    \cite{BPP} \\  
    - & 83(12) & - & - & - & 80(40) & \cite{KN_PRD_02} \\ 
    0(10) & 114(10) & 22(5) & - & 0 & 136(25) & \cite{MV_04} \\ 
    - & - & - & - & - & 110(40) & \cite{HLbL_2007} \\ 
    $-19(19)$ & 114(13) & 15(10) & $-7(7)$ & 2.3 [c-quark] & 105(26) & 
    \cite{PdeRV_09} \\ 
    $-19(13)$ & 99(16) & 22(5) & $-7(2)$ & 21(3) & 116(40) &
    \cite{N_09,JN_09} \\
    - & 81(2) & - & - & 107(2) & 188(4) & \cite{FGW} \\
    - & -     & - & - & -      & 118-148 & \cite{BM} \\
    - & 68(3) [$\pi^0$] & - & - &  82(6) & 150(3) & \cite{GdeR} \\
    - & -     & - & - & -      & 76(4)-125(7) &
    \cite{Masjuan_Vanderhaeghen} \\ 
    $-(11-71)$ & - & - & - & - & - & \cite{EPR-M} \\ 
    $-20(5)$   & - & - & - & - & - & \cite{BZ-A} \\ 
    $-45$ & $+\infty$ & - & - & $60$ & - & no FF  \\ 
    \hline 
  \end{tabular}
  \end{center} 
\end{table}

\begin{figure}[h]
\centerline{\includegraphics[width=0.8\textwidth]{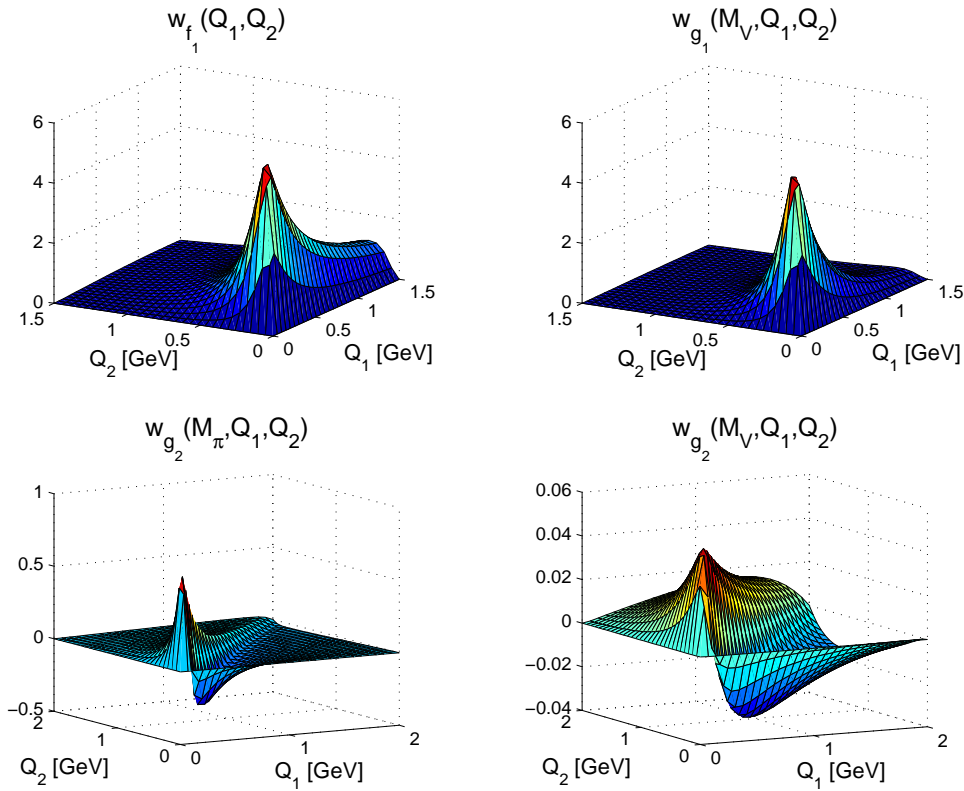}}
\caption{The different weight functions $w_i$ in 
  Eq.~(\ref{2D_integral}) from Ref.~\cite{KN_PRD_02}.}  
\label{fig:2D_weightfunctions}
\end{figure}

A selection of estimates for HLbL is presented in
Table~\ref{tab:HLbL_summary}. Note that the Refs.~\cite{HKS,BPP} are
the only full calculations of HLbL to date, using, as much as
possible, one model for all the contributions (HLS model in
Ref.~\cite{HKS}, Extended Nambu-Jona-Lasinio (ENJL) model in
Ref.~\cite{BPP}). Both calculations showed that the exchanges of the
lightest pseudoscalar states, $\pi^0,\eta,\eta^\prime$, dominate
numerically, which can be understood from the large-$N_c$
counting. The contributions from the (dressed) pion-loop and the
(dressed) quark-loop are subdominant, but not negligible, and they
happen to largely cancel each other numerically. The final results for
the total HLbL contribution were rather close in both models. In
Ref.~\cite{KN_EPJC_01} an ansatz for the pion-photon transition form
factor with a minimal number of narrow vector resonances in
large-$N_c$ QCD (lowest meson dominance (LMD, LMD+V)) was matched to
short-distance constraints from the OPE. The reevaluation of the
pion-pole contribution to HLbL in Ref.~\cite{KN_PRD_02} with the
ansatz from Ref.~\cite{KN_EPJC_01} then revealed a sign error in the
earlier calculations~\cite{HKS,BPP}. Furthermore, for the neutral pion
contribution, a 2-dimensional integral representation was derived in
Ref.~\cite{KN_PRD_02} for a certain class of vector meson dominance
(VMD)-like form factors. In the end, one integrates over the lengths
of the two independent loop-momenta $|Q_1|,|Q_2|$ in Euclidean
space. Schematically,
\begin{equation} \label{2D_integral}
a_{\mu}^{{\rm{HLbL}};\pi^0} = \int_0^\infty dQ_1
\int_0^\infty dQ_2 \, \sum_i w_i(Q_1, Q_2) \, f_i(Q_1, Q_2), 
\end{equation} 
with universal weight functions $w_i$ and where the (model) dependence
on the form factors resides in the functions $f_i$, see
Ref.~\cite{KN_PRD_02} for details. Note that the expressions in
Ref.~\cite{KN_PRD_02} with on-shell form factors are in general not
valid as they stand. As pointed out in Ref.~\cite{MV_04}, one needs to
set the form factor at the external vertex to a constant to obtain the
pion-{\it pole\ } contribution. But the expressions are valid for the
constant Wess-Zumino-Witten (WZW) form factor and the off-shell VMD
form factor. The weight functions $w_i$ are plotted in
Figure~\ref{fig:2D_weightfunctions}. One can see that the relevant
momentum regions are about $0 - 1.25~\mbox{GeV}$, see also the
numerical analysis in Ref.~\cite{CD12_Proceedings}.  As long as the
form factors in different models lead to a damping, we expect
comparable results for $a_{\mu}^{{\rm{HLbL}};\pi^0}$ at the level of
20\%, which is indeed the case, see the papers quoted in
Table~\ref{tab:HLbL_summary} and Ref.~\cite{other_recent_papers}.  For
the general case with off-shell form factors, a 3-dimensional integral
representation was derived in Ref.~\cite{JN_09}, where the
integrations run over $Q_1, Q_2$ and $\cos\theta$, where $Q_1 \cdot
Q_2 = |Q_1| |Q_2| \cos\theta$.

Later, Ref.~\cite{MV_04} derived new short-distance constraints from
the four-point function on the pion-pole and axial-vector-pole
contributions, which do not allow for any form factors at the external
vertex. Reference~\cite{MV_04} also included the mixing of two
axial-vector nonets and studied the pion-loop within the HLS model in
more detail. All this lead to a substantial enhancement of these
contributions to HLbL. More recently, a new short-distance constraint
on the off-shell form factor at the external vertex in pion-exchange
was derived in Ref.~\cite{N_09}, which yielded, again in the framework
of the lowest meson dominance approximation to large-$N_c$ QCD, a
value for this contribution about half-way between the results in
Refs.~\cite{HKS,BPP,KN_PRD_02} and those in Ref.~\cite{MV_04}. Note
that the compilations~\cite{HLbL_2007,PdeRV_09,JN_09} and
Ref.~\cite{N_09} are largely based on the full
calculations~\cite{HKS,BPP}, with revised or newly calculated values
for some of the contributions. More recent estimates, mostly for the
pseudoscalar contribution, can be found in
Ref.~\cite{other_recent_papers}. While most of these evaluations agree
at the level of 15\%, if one takes the extreme values, there is a
spread of $a_\mu^{\rm HLbL; PS} = (59 - 107) \times 10^{-11}$.

Until 2010, a consensus had been reached about the central value
$a_\mu^{\rm HLbL} \approx 110 \times 10^{-11}$, but there was a
discussion on how to estimate the error, more progressively, $\pm 26
\times 10^{-11}$, in Ref.~\cite{PdeRV_09} and more conservatively,
$\pm 40 \times 10^{-11}$, in Refs.~\cite{N_09,JN_09}. In view of the
precision goal of future $g-2$ experiments at Fermilab and
J-PARC~\cite{Fermilab_J-PARC} with $\delta a_\mu = 16 \times 10^{-11}$
and the continued progress in improving the error in HVP, the HLbL
contribution might soon be the main uncertainty in the theory
prediction, if it cannot be brought under better
control~\cite{JN_09,Miller_etal_12,Venanzoni_talk_Blum_et_al_13}.

In the last few years, several works have appeared which yield much
larger (absolute) values for some of the contributions, see
Table~\ref{tab:HLbL_summary}. In Ref.~\cite{FGW} the quark-loop was
studied using a Dyson-Schwinger equation approach. In contrast to
Refs.~\cite{HKS,BPP}, no damping compared to the bare constituent
quark-loop result was seen, when a dressing was included. Note that
this calculation of the quark-loop is not yet complete and that
earlier, very large results for the quark-loop seem to have been
affected by some errors in the numerics in certain parts. The large
size of the quark-loop contribution in Ref.~\cite{FGW} was questioned
in the papers~\cite{BM,GdeR}, using different quark-models and
approaches, see also the ballpark prediction for HLbL in
Ref.~\cite{Masjuan_Vanderhaeghen}.  The pion-loop contribution was
analyzed in Ref.~\cite{EPR-M}. The authors stressed the importance of
the pion-polarizability effect and the role of the axial-vector
resonance $a_1$, which are not included in the models used in
Refs.~\cite{HKS,BPP}. Depending on the value of the
pion-polarizability and the model for the $a_1$ resonance used, a
large variation was seen. The issue was taken up in Ref.~\cite{BZ-A}
where different models for the pion-loop were studied. The inclusion
of the $a_1$ resonance was attempted, but no finite result for $g-2$
could be achieved. With a cutoff of $1~\mbox{GeV}$, a result close to
the earlier estimate in Ref.~\cite{BPP} was
obtained. Reference~\cite{BZ-A} also pointed out that the very small
(absolute) value for the pion-loop in Ref.~\cite{HKS} could be due to
the fact that the HLS model used in Ref.~\cite{HKS} has a wrong
high-energy behavior and that there is some cancellation between
positive and negative contributions in the pion-loop in HLbL.

\section{Outlook}

Concerning the future, maybe lattice QCD will provide a reliable
calculation of HLbL at some point, see Ref.~\cite{HLbL_Lattice} for
some promising recent results. In the meantime, only a close
collaboration between theory and experiment can lead to a better
controlled estimate for HLbL. On the theory side, the hadronic models
can be improved by short-distance constraints from perturbative QCD to
have a better matching at high momenta. One can also use dispersion
relations to connect the theory with experimental data, e.g. in
$\gamma\gamma \to \pi\pi$~\cite{DR_approach}.  Also the issue about
whether the dressing of the bare constituent quark-loop leads to a
suppression or an enhancement needs to be studied further. This
problem is also related to the question whether there is any
double counting involved. 

On the experimental side, the information on various processes
(decays, form factors, cross-sections) of hadrons interacting with
photons at low and intermediate momenta, $|q|~\leq~2~\mbox{GeV}$, can
help to constrain the models. Important experiments which should be
pursued include more precise measurements of the (transition) form
factors of light pseudoscalars with possibly two off-shell photons in
the process $e^+ e^- \to e^+ e^- P \ (P= \pi^0, \eta, \eta^\prime)$
and the two-photon decay width and the (double) Dalitz decays of these
mesons. This could further reduce the error of the dominant
pseudoscalar exchange contribution~\cite{KLOE-2_impact}. Concerning
the pion-loop contribution, in addition to studying $\gamma\gamma \to
\pi\pi$, measurements of the pion-polarizability in various processes,
e.g.\ in radiative pion decay $\pi^+ \to e^+ \nu_e \gamma$, in
radiative pion photoproduction $\gamma p \to \gamma^\prime \pi^+ n$ or
with the hadronic Primakoff effect $\pi A \to \pi^\prime \gamma A$ or
$\gamma A \to \pi^+ \pi^- A$ (with some nucleus $A$), can help to
improve the models~\cite{EPR-M}. For the development of models with
the axial-vector resonance $a_1$ and estimates of the sizable
axial-vector contribution, information about the decays $a_1 \to
\rho\pi, \pi\gamma$ would be useful as well. Finally, to extract the
needed quantities from experiment will also require the development of
dedicated Monte Carlo programs for the relevant
processes~\cite{RMC_WG}.

\section{Conclusions}

If the recent results for the quark-loop and pion-loop are taken at
face value, one obtains the range $a_\mu^{\rm HLbL} = (64 - 202)
\times 10^{-11}$. While the new approaches raise some important issues
and point to potential shortcomings in the previously used models,
these calculations are also still preliminary and further studies are
needed. Therefore, the estimate
\begin{equation}
a_\mu^{\rm HLbL} = (116 \pm 40) \times 10^{-11} 
\end{equation}
from Refs.~\cite{N_09,JN_09} still seems to give a fair description of
the current situation.

\section*{Acknowledgments} 

I would like to thank the organizers of the {\it International
  Workshop on $e^+e^-$ collisions from $\Phi$ to $\Psi$ 2013
  (PHIPSI13)}, of the {\it Working Group on Radiative Corrections and
  Monte Carlo Generators for Low Energies} and of the {\it LC13
  Workshop: Exploring QCD from the infrared regime to heavy flavour
  scales at B-factories, the LHC and a Linear Collider} for the
opportunity to present reviews on the HLbL in the muon $g-2$. Generous
support from the organizers, the INFN, the ECT$\star$, Trento, Italy,
and the Heinrich-Greinacher-Stiftung, Physics Institute, University of
Bern, Switzerland, is gratefully acknowledged. The AEC is supported by
the ``Innovations- und Kooperationsprojekt C-13'' of the
``Schweizerische Universit\"atskonferenz SUK/CRUS.''

\end{document}